%% file: kinematics_reconstruction.tex
\documentclass[a4paper,11pt]{article}
\usepackage{jheppub}
\usepackage[T1]{fontenc}
\usepackage{microtype}
\usepackage{dcolumn}
\usepackage{booktabs}
\usepackage{chemformula}
\usepackage{multirow}
\usepackage{comment}
\usepackage{lineno}  % for line numbering during review
\usepackage{xspace}

\def\bfseries{\fontseries \bfdefault \selectfont \boldmath}

\input{src/commands.tex}

\arxivnumber{2502.10198} % if you have one

\title{Reconstructing neutrinoless double beta decay event kinematics in a xenon gas detector with vertex tagging}

\input{src/authors.tex}

\abstract{
If neutrinoless double beta decay is discovered, the next natural step would be understanding the lepton number violating physics responsible for it. Several alternatives exist beyond the exchange of light neutrinos. Some of these mechanisms can be distinguished by measuring phase-space observables, namely the opening angle $\cos\theta$ among the two decay electrons, and the electron energy spectra, $T_1$ and $T_2$. In this work, we study the statistical accuracy and precision in measuring these kinematic observables in a future xenon gas detector with the added capability to precisely locate the decay vertex. For realistic detector conditions (a gas pressure of 10~bar and spatial resolution of 4~mm), we find that the average $\overline{\cos\theta}$ and $\overline{T_1}$ values can be reconstructed with a precision of 0.19 and 110~keV, respectively, assuming that only 10 neutrinoless double beta decay events are detected. 
}

\begin{document}
 
\maketitle
\flushbottom

\input{src/intro.tex}
\input{src/detector.tex}

\input{src/methods.tex}
\input{src/reconstruction_performance.tex}

\input{src/statistical_analysis.tex}

\input{src/conclusions.tex}

\acknowledgments
\input{src/ack.tex}

\bibliographystyle{JHEP}
\bibliography{biblio}

\end{document}

%% file: src/commands.tex
\newcommand{\bbnonu}{\ensuremath{0\nu\beta\beta}\xspace}

\newcommand{\Ba}[1]{\ensuremath{^{#1}\mathrm{Ba}}}
\newcommand{\Xe}[1]{\ensuremath{^{#1}\mathrm{Xe}}}

\newcommand{\Kr}[1]{\ensuremath{^{#1}\mathrm{Kr}}\xspace}

\newcommand{\mbb}{\ensuremath{\langle m_{\beta\beta} \rangle}}
\newcommand{\halflife}{\ensuremath{T_{1/2}}\xspace}

\newcommand{\nexus}{\texttt{nexus}\xspace}
\newcommand{\nudobe}{$\nu$\texttt{DoBe}\xspace}

%% file: src/authors.tex
\collaboration{The NEXT Collaboration}

\author[22,17]{M.~Mart\'inez-Vara,}
\author[3]{K.~Mistry,}
\author[17,a]{F.~Pompa\note[a]{Non-NEXT author. Now at SUBATECH, France.},}
\author[3]{B.J.P.~Jones,}
\author[17]{J.~Mart\'in-Albo,}
\author[17]{M.~Sorel,}
\author[2]{C.~Adams,}
\author[16]{H.~Almaz\'an,}
\author[24]{V.~\'Alvarez,}
\author[20]{B.~Aparicio,}
% ...
\affiliation[2]{
Argonne National Laboratory, Argonne, IL 60439, USA}
\affiliation[3]{
Department of Physics, University of Texas at Arlington, Arlington, TX 76019, USA}
% ...
\affiliation[16]{
Department of Physics and Astronomy, Manchester University, Manchester. M13 9PL, United Kingdom}
\affiliation[17]{
Instituto de F\'isica Corpuscular (IFIC), CSIC \& Universitat de Val\`encia, Calle Catedr\'atico Jos\'e Beltr\'an, 2, Paterna, E-46980, Spain}
% ...
\affiliation[20]{
Department of Organic Chemistry I, Universidad del Pais Vasco (UPV/EHU), Centro de Innovaci\'on en Qu\'imica Avanzada (ORFEO-CINQA), San Sebasti\'an / Donostia, E-20018, Spain}
% ...
\affiliation[22]{
Donostia International Physics Center, BERC Basque Excellence Research Centre, Manuel de Lardizabal 4, San Sebasti\'an / Donostia, E-20018, Spain}
% ...
\affiliation[24]{
Instituto de Instrumentaci\'on para Imagen Molecular (I3M), Centro Mixto CSIC - Universitat Polit\`ecnica de Val\`encia, Camino de Vera s/n, Valencia, E-46022, Spain}
\emailAdd{mavam@ific.uv.es}
\emailAdd{krishan.mistry@uta.edu}
\emailAdd{federica.pompa@subatech.in2p3.fr}

%% file: src/intro.tex
\section{Introduction}
\label{sec:intro}
Neutrinoless double beta (\bbnonu) decay is a hypothetical radioactive process in which two neutrons in a nucleus simultaneously transform into two protons, emitting two electrons: \ch{^{A}_{Z}X $\to$ ^{A}_{Z+2}X$'$ + 2 e-}. Such a transition would violate the conservation of total lepton number by two units ($\Delta L=2$), and hence it is forbidden in the Standard Model (SM) of particle physics. The search for \bbnonu decay is, in fact, the most sensitive known test of \emph{lepton number violation} (LNV). At present, the best experimental limits on the half-life of the decay exceed $10^{26}$~years \cite{GERDA:2020xhi, KamLAND-Zen:2024eml}.\footnote{For an in-depth discussion of both the theoretical and experimental aspects of \bbnonu decay, we refer the reader to two recent review papers on the topic \cite{Agostini:2022zub, Gomez-Cadenas:2023vca}.}  

Most often, \bbnonu decay is described as taking place through the so-called \emph{mass mechanism}, that is, the exchange of light, active Majorana neutrinos via SM left-handed weak currents. However, other LNV mechanisms ---\thinspace arising, in general, in extensions of the SM involving new physics at energy scales above the electroweak one\thinspace--- could also mediate the decay and contribute to its rate.\footnote{See, for instance, ref.~\cite{Rodejohann:2011mu} for a comprehensive list and discussion of these alternative \bbnonu-decay mediating mechanisms.} These various contributions would have to be disentangled, in the case of a \bbnonu-decay discovery, to conclude anything on the features of the LNV sources and, in particular, on the scale of neutrino masses. There are several ways to do that, including a comparison of the \bbnonu-decay rate in two or more isotopes (e.g. $^{136}$Xe and $^{100}$Mo or $^{76}$Ge and $^{100}$Mo) \cite{Agostini:2022bjh} or the decay to excited states \cite{Simkovic:2001qf}, the measurement of the angular distribution of the \bbnonu electrons \cite{Doi:1983wv, Ali:2007ec, SuperNEMO:2010wnd}, or correlating the \bbnonu-decay rate with other probes such as cosmological observations \cite{Deppisch:2004kn} or experiments searching for heavy neutral leptons \cite{Hernandez:2022ivz}.

Here, we evaluate the ability of the NEXT technology to discriminate among different LNV mechanisms contributing to the decay amplitude via the reconstruction of the electron tracks in the detector. The NEXT detector is a high-pressure xenon gas time projection chamber that searches for \bbnonu-decay: $^{136}\textnormal{Xe} \to ^{136}\textnormal{Ba}^{2+}+2\textnormal{e}^-$. The detector can produce high-resolution images of large tracks down to mm resolution with excellent energy resolution. The distributions of the kinematic observables of opening angle and single-electron kinetic energy can be reconstructed if the decay vertex location within the detector volume is determined. This can be achieved either through the identification (or \emph{tagging}) of the daughter barium isotope, a technique currently under development in NEXT, or through advanced image analysis methods.

%%%%%%%%%%
\begin{figure}
\centering
\includegraphics[width=0.65\linewidth]{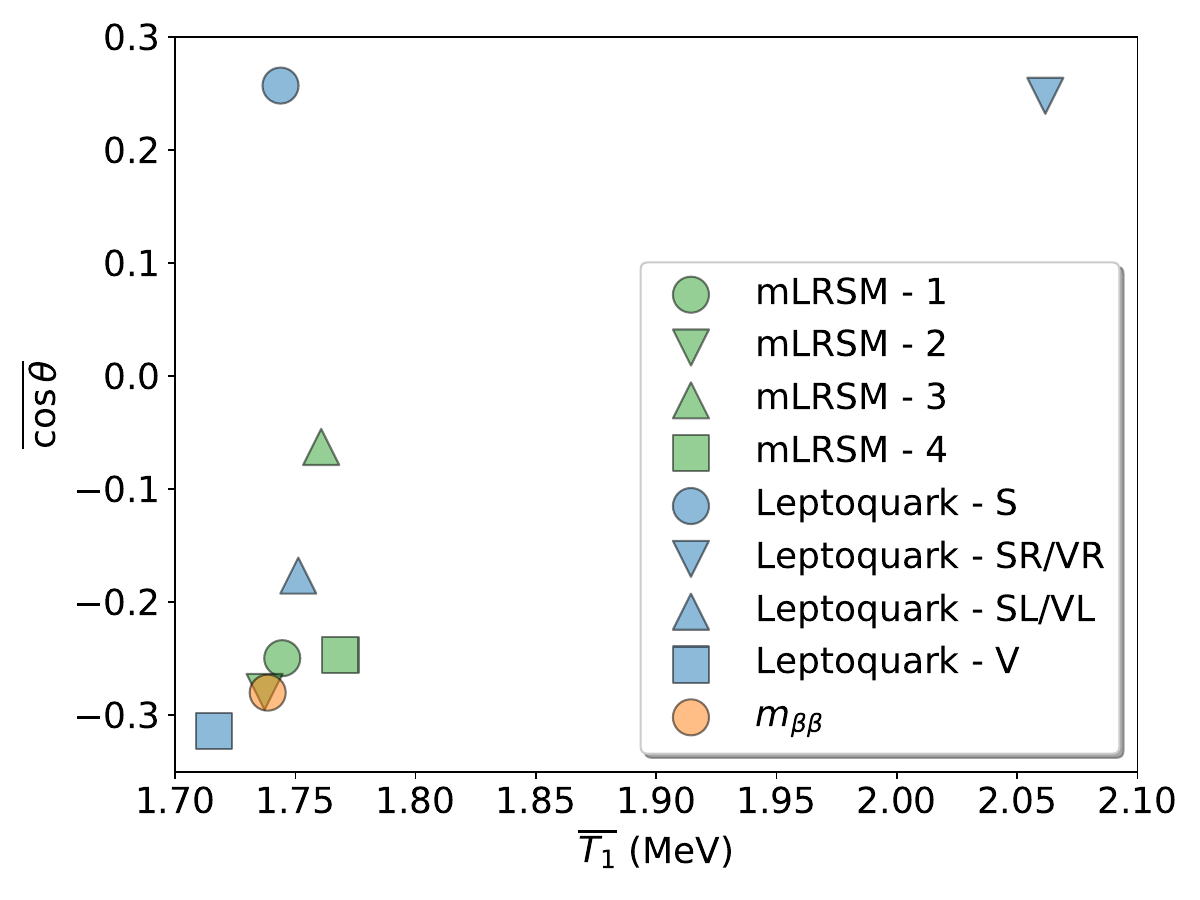}
\caption{Average cosine of the opening angle between decay electrons versus the average kinetic energy of the most energetic electron in the decay, for \Xe{136} and some benchmark \bbnonu-decay models: standard neutrino mass mechanism ($m_{\beta\beta}$), four different realizations of the minimal left-right symmetric Standard Model (mLRSM models in~\cite{Scholer:2023bnn}), and four different leptoquark contributions (leptoquark models in~\cite{Graf:2022lhj}).}
\label{fig:model_kinematics}
\end{figure}
%%%%%%%%%%

Figure~\ref{fig:model_kinematics} shows example particle physics models \cite{Scholer:2023bnn,Graf:2022lhj} readily available within the \texttt{$
\nu$DoBe} package~\cite{Scholer:2023bnn}, the model generator we consider in this work. The models are shown in terms of the average cosine of the opening angle between decay electrons ($\overline{\cos\theta}$) and the average kinetic energy of the most energetic electron ($\overline{T_1}$)\footnote{To avoid any ambiguities in the 1$\leftrightarrow$2 electron labeling, throughout this work we assume that $T_{1} > T_{2}$ in each decay, implying $T_{1}>Q/2$.} considering an ideal experiment detecting with perfect event reconstruction capabilities a sufficiently large number of \Xe{136} \bbnonu events to result in negligible statistical uncertainties, and adopting realistic nuclear and atomic physics inputs to compute the phase-space observables. For the standard mass mechanism, we expect $\overline{T_1}=1.74$~MeV and $\overline{\cos\theta}=-0.28$ in \Xe{136}. Despite the fact that all models span the entire phase-space available ($Q/2<T_1<Q$ and $-1<\cos\theta<1$, where $Q=2.458$~MeV~\cite{Redshaw:2007un} is the \Xe{136} \bbnonu decay $Q$ value), the kinematic observables averaged over the signal datasets feature potential model discrimination sensitivity. As the figure shows, deviations from the standard neutrino mass predictions as large as $\sim$300~keV and $\sim$0.5 are expected for $\overline{T_1}$ and $\overline{\cos\theta}$, respectively. In this work, we specifically try to assess the uncertainties on these kinematic observables arising from finite signal sample statistics, from realistic event kinematic reconstruction assumptions, and from current nuclear and atomic physics uncertainties. 

The structure of the paper is the following: section~\ref{sec:detector} provides a concise overview of the NEXT project and technology, including the barium tagging technique; section~\ref{sec:methods} describes the methodologies used in this study to simulate and reconstruct \bbnonu event kinematics; sections~\ref{sec:reconstruction_performance} and \ref{sec:statistical_analysis} present the findings of the study, in terms of reconstruction performance for event-based and dataset-based kinematic observables, respectively; lastly, section~\ref{sec:conclusions} offers concluding remarks. 

%% file: src/detector.tex
\section{Detector}
\label{sec:detector}
The \emph{Neutrino Experiment with a Xenon TPC} (NEXT) is an international effort dedicated to the search for \bbnonu decay in \Xe{136} using high-pressure xenon gas time projection chambers (HPXeTPC) with amplification of the ionization signal by electroluminescence (EL). This detector technology takes advantage of the inherently low fluctuations in the production of ionization pairs (i.e., small Fano factor) in xenon gas to achieve an energy resolution significantly better than that of other \Xe{136}-based double-beta decay experiments \cite{Nygren:2009zz}. Moreover, the tracks left in gaseous xenon by \bbnonu events have distinct features that can be used for background rejection~\cite{NEXT:2016ire,NEXT:2019gtz,NEXT:2020jmz,NEXT:2021vzd} and the gas pressure can be varied to explore different track sizes. 

Over the last decade, the NEXT Collaboration has proven the performance of the HPXeTPC technology in the key parameters required for the observation of \bbnonu decay. The NEXT concept was initially tested in small, surface-operated detectors \cite{NEXT:2012lrw, NEXT:2012rto, NEXT:2013kkl, NEXT:2014yrq, NEXT:2015rel}. This phase was followed by the underground operation at the Laboratorio Subterr\'aneo de Canfranc (LSC) of NEXT-White \cite{NEXT:2018rgj}, a radiopure HPXeTPC containing approximately 5~kg of xenon at 10~bar pressure. The results obtained with NEXT-White include the development of a procedure to calibrate the detector using \Kr{83\mathrm{m}} decays \cite{NEXT:2018sqd}, measurement of an energy resolution better than 1\% FWHM at 2.5~MeV \cite{NEXT:2018keh, NEXT:2019qbo}, demonstration of robust discrimination between single-electron and double-electron tracks \cite{NEXT:2019gtz, NEXT:2020jmz, NEXT:2021vzd}, measurement of the radiogenic background, validating the accuracy of our background model \cite{NEXT:2018zho, NEXT:2019rum},  measurement of the half-life of the standard double beta decay of \Xe{136} via direct background subtraction \cite{NEXT:2021dqj}, and a search for \bbnonu decay in \Xe{136} \cite{NEXT:2023daz}.

The NEXT-100 detector \cite{NEXT:2015wlq}, which started operations at LSC in the first half of 2024, constitutes the third phase of the program. It is an HPXeTPC containing about 100~kg of xenon (enriched at 90\% in \Xe{136}) at 15~bar pressure. In addition to a physics potential competitive with the best current experiments in the field, NEXT-100 can be considered as a large scale demonstrator of the suitability of the HPXeTPC technology for detector masses at the tonne scale. The detector will reach a sensitivity to \bbnonu decay exceeding $10^{25}$~yr at 90\% confidence level after a run of 3~years.

The HPXeTPC technology can be scaled up to multi-tonne source masses by introducing several already available technological advancements \cite{NEXT:2020amj}. A future tonne-scale NEXT module will be able to improve by more than one order of magnitude the current half-life limit in \Xe{136}. Higher definition tracking modalities~\cite{NEXT:2023mou,NEXT:2024btf} could reduce readout pixelization areas towards 3-4~mm$^2$ while electron diffusion could be reduced to levels of 1.5~mm/$\sqrt m$ achieved by addition of percent-level molecular dopants~\cite{Azevedo:2015eok, Felkai:2017oeq, NEXT:2019doo}. Improved image analysis approaches such as image de-convolution~\cite{NEXT:2021vzd} and machine learning~\cite{NEXT:2016ire, NEXT:2020jmz}  are also expected to continue to improve the power of NEXT topological analyses in future phases.  As such, additional information could potentially become extractable from the NEXT tracks, including the decay vertex location, opening angle, and individual kinetic energy of the decay electrons. The physics content of the information encoded in the latter observables is the subject of this paper.

The HPXeTPC technology may permit even a more radical approach to the next generation of \bbnonu-decay experiments by implementing a system capable of detecting with high efficiency the presence of the \Ba{^{2+}} ion produced in the \Xe{136} \bbnonu decay. The detection would occur in (delayed) coincidence with the identification of the two electrons, offering the potential for a background-free measurement from non-$\beta\beta$ sources. The $2\nu\beta\beta$ decay would still remain as an irreducible background, although its contribution can be made negligible with good energy resolution. To realize this technology, the NEXT Collaboration is exploring the use of Single Molecular Fluorescence Imaging (SMFI) to identify \Ba{^{2+}} ions. SMFI employs molecules that include both fluorescent and metal-binding groups that inhibit fluorescence unless the molecules are chelated with a suitable ion such as barium. The possibility of using SMFI as the basis of molecular sensors for barium tagging was proposed in ref.~\cite{Jones:2016qiq}, followed, shortly after, by a proof of concept which managed to resolve individual \Ba{^{2+}} ions on a scanning surface using an SMFI-based sensor \cite{McDonald:2017izm}. Rapid advances have been made for this purpose subsequently~\cite{NEXT:2023qoy, NEXT:2022ita, Thapa:2019zjk, Thapa:2020gjw, jones2022barium, thapa2021demonstration, McDonald:2017izm, Rivilla:2020cvm, Bainglass:2018odn, NEXT:2023mou, Jones:2016qiq, NEXT:2021idl}. A barium-tagging NEXT detector with a mass in the tonne range could reach a sensitivity exceeding 10$^{28}$~years~\cite{Giuliani:2019uno}. 

While a full barium-tagging detector design for a NEXT tonne scale detector has yet to be finalized, there exist potential schemes capable of localizing the decay vertex with superior position resolution compared to the electron-sensing TPC detector. To extract decay vertex information, a fluorescence imaging sensor has to be able to image the area around the \Ba{^{2+}} ion after it has drifted to the TPC cathode. The diffusion of barium ions over large drifts ($\sim$m) is orders of magnitude smaller than the diffusion of ionization electrons due to their higher mass, introducing negligible position smearing. Given the limited field of view of the barium tagging sensors, fluorescence imaging requires controlling either the position of the sensor \cite{Rivilla:2020cvm} or the position of the ion \cite{Bainglass:2018odn, NEXT:2021idl} over the entire cathode plane. 

The precision of vertex information in realistic detector conditions and its impact on the extracted kinematic variables using traditional TPC reconstruction and/or barium tagging will be the subject of future work. However, given vertex information made accessible, an important question remains: to what extent do the resulting events ---\thinspace with tracks that follow complex trajectories due to multiple scattering, stochastic energy losses, and delta ray emissions, among other effects\thinspace--- preserve information about the underlying physics of the \bbnonu-decay mechanism? This is the central topic of investigation of the present work. For concreteness, we assume that the decay vertex position can be reconstructed in an idealized detector with barium tagging with negligible (sub-mm) position smearing, regardless of the TPC readout granularity, and explore the capability of voxelized tracking algorithms to infer information about the underlying LNV processes that drive neutrinoless double beta decay.

%% file: src/methods.tex
\section{Methods}
\label{sec:methods}
In order to assess the power of a tonne-scale xenon-gas detector with barium-tagging capabilities to discriminate the contribution among different \bbnonu operators, we proceed in three steps which include: event generation, detector simulation, and event reconstruction.

\subsection{Event generation}
Neutrinoless double beta decay event generation is performed via \nudobe \cite{Scholer:2023bnn}. This tool can be used for automated calculations of \bbnonu~rates and electron kinematics for all isotopes of experimental interest, for LNV operators up to and including dimension 9. A given LNV model can be specified in terms of Wilson coefficients of either \emph{Standard Model Effective Field Theory} (SMEFT) or \emph{Low-energy Effective Field Theory} (LEFT) relevant operators, above or at the electroweak scale, respectively. In case of SMEFT, \nudobe automatically performs operator matching at the electroweak scale. To compute \bbnonu~rates and kinematics, it is possible to choose among various sets of nuclear matrix elements (NMEs), based on different many-body nuclear-structure methods, and among two different approximation schemes for the computation of the electron phase-space factors (PSFs). Short-range NME contributions associated to hard-neutrino-exchange processes can also be accounted for through hadronic low-energy constants. 

In our case, we perform our analysis in terms of LEFT operators to generate electron kinematics in \Xe{136}, particularly single-electron kinetic energy spectra and angular correlations among the two decay electrons. Unless otherwise noted, we generate events according to the standard neutrino mass mechanism, driven by the dimension-3 operator $\mathcal{O}_{m_{\beta\beta}}$ and labeled as {\mbb} in the following. For our nominal analysis, we adopt a shell-model calculation \cite{Agostini:2022zub,Menendez:2017fdf} for the \Xe{136} NMEs, and PSFs calculated from approximate wave functions in a uniform charge distribution (so-called PSF ``scheme A" in ref.~\cite{Scholer:2023bnn}). We also neglect unknown low-energy constants that would induce short-range contributions to \bbnonu NMEs, such as the one induced by the $g_{\nu}^{NN}$ operator (see \cite{Scholer:2023bnn} for details). In section~\ref{sec:statistical_analysis}, we study the effect on electron kinematics from nuclear, hadronic and atomic physics assumptions that depart from the nominal ones outlined above. 

%%%%%%%%%%
\begin{figure}
\centering
\includegraphics[width=\linewidth]{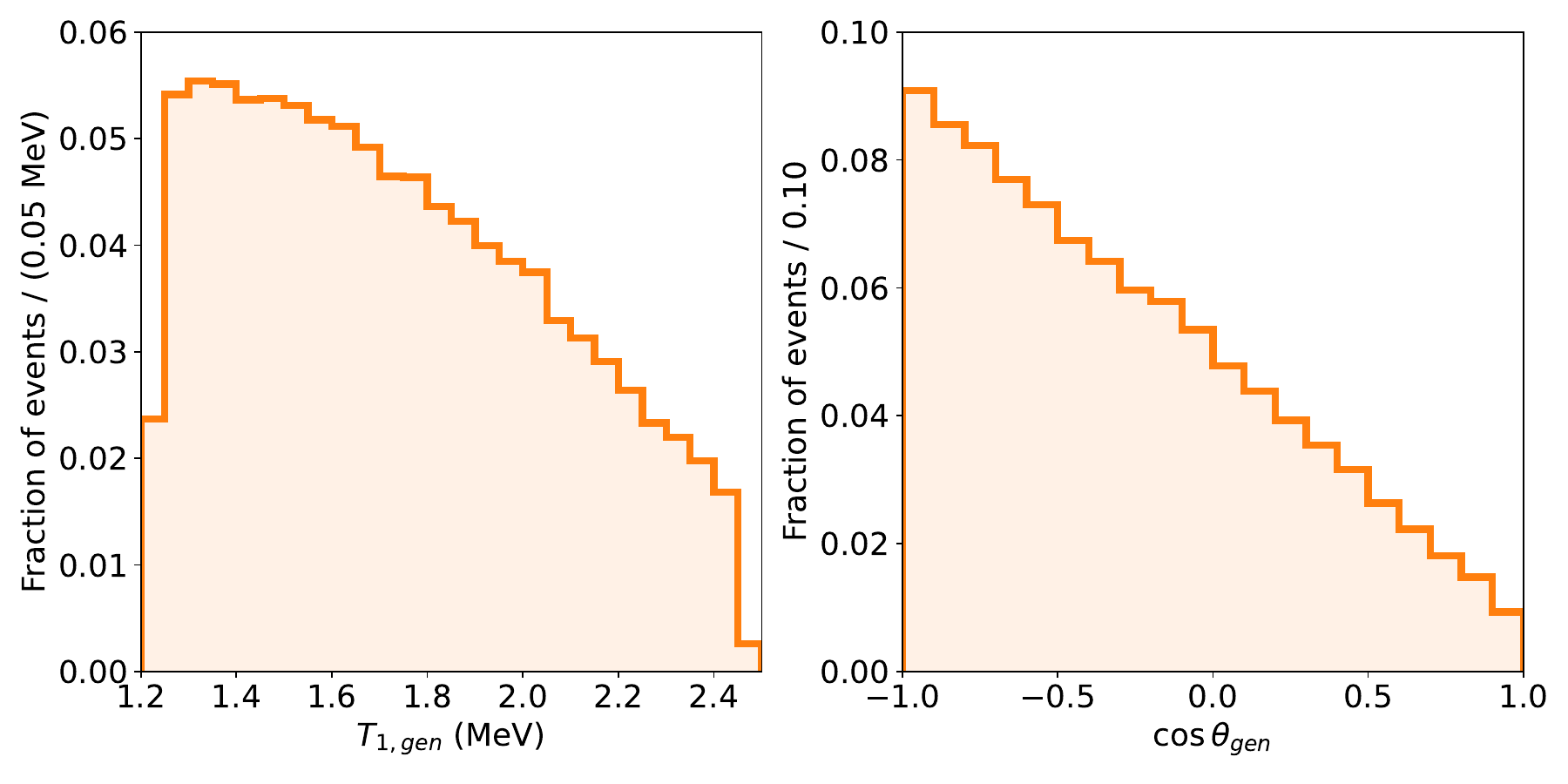}
\caption{Generator-level kinematic quantities for the standard neutrino mass \bbnonu-decay model from \nudobe. Left, generated single-electron kinetic energy, $T_{1,\mathrm{gen}}$. Right, generated cosine of the opening angle between the two decay electrons, $\cos\theta_\mathrm{gen}$. Electrons are labelled such that $T_{1,\mathrm{gen}}>T_{2,\mathrm{gen}}$.
\label{fig:event_generation}}
\end{figure}
%%%%%%%%%%

Figure~\ref{fig:event_generation} shows the distributions of the single-electron kinetic energy ($T_{1,\mathrm{gen}}$) and opening angle between the emitted electrons ($\cos\theta_\mathrm{gen}$) for \Xe{136} \bbnonu decay driven by the {\mbb}~contribution. To avoid any confusion in our notation, the subscript `gen' has been added here to indicate that these are generator-level observables, which do not account for any detector or reconstruction effects. Concerning $T_{1,\mathrm{gen}}$, the distribution peaks at the lowest possible value of $Q/2$, inducing a relatively low value of $\overline{T_1}$ for the \mbb~contribution in fig.~\ref{fig:model_kinematics}, and a relatively small asymmetry in the energy spectra of the most energetic ($e_1$) and the least energetic ($e_2$) electron in the decay. For what concerns the angular correlation, the two electrons are preferentially emitted back to back ($\cos\theta_\mathrm{gen}=-1$) for the {\mbb} model.

\subsection{Detector simulation}
The propagation of the \bbnonu electrons in gaseous xenon is simulated via \nexus \cite{Martin-Albo:2015dza}, the Geant4-based simulation framework \cite{Allison:2016lfl, Allison:2006ve, GEANT4:2002zbu} of the NEXT Collaboration, and Geant4 release 11.2.1. For each event, the initial kinematics of the primary electrons is provided by \nudobe. Xenon gas at either 1, 5, 10 or 15~bar is used as sensitive volume, to assess how our results depend on gas pressure. A total of 6,000 simulated events are processed for each detector pressure setting. For simplicity, no detector border effects are considered; a sufficiently large active volume around the \bbnonu-decay vertex positions is simulated to ensure full event containment. 

\begin{figure}
\centering
 \includegraphics[width = 0.65
\textwidth]{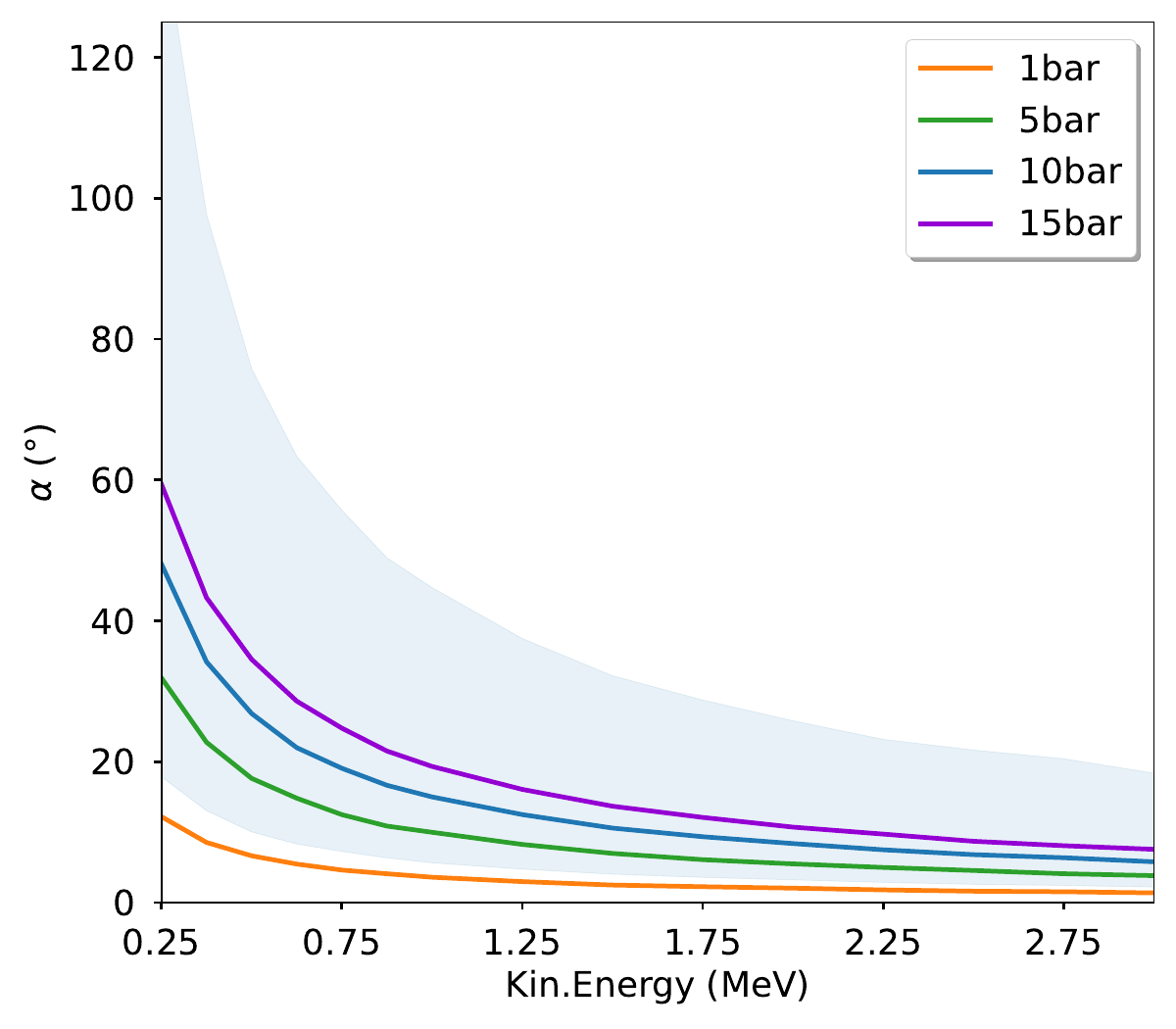}
\caption{Mean scattering angle $\alpha$ for electrons after going through a layer of 1~mm Xe gas for different gas densities: $0.56 \times 10^{-3}~\mathrm{g/cm^{3}}$ (1~bar), $2.8 \times 10^{-2}~\mathrm{g/cm^{3}}$ (5~bar), $5.6 \times 10^{-2}~\mathrm{g/cm^{3}}$ (10~bar) and $8.4 \times 10^{-2}~\mathrm{g/cm^{3}}$ (15~bar) at 291~K. The shaded band represents the standard deviation of $\alpha$ at 10~bar.}
\label{fig:ang_Ek}  
\end{figure}

All relevant electromagnetic physics processes involving electrons, positrons and photons in the keV--MeV energy range are considered. Of particular importance is the simulation of elastic scattering affecting the primary electrons. The Geant4 physics list employed, \texttt{G4EmStandardPhysics\_option4}, described in the documentation as the most accurate set of models of electromagnetic physics, combines a Goudsmit-Sounderson multiple Coulomb scattering (MCS) model \cite{Novak:2024} to describe continuous, small-angle scattering with a single Coulomb scattering model applied for large angle scattering. The simulation is performed with a maximum step size among Geant4 hits of 4~mm. A 1~mm maximum step size was also studied but gave negligible differences. Figure~\ref{fig:ang_Ek} presents the simulated mean  scattering angle ($\alpha$) of an electron after traversing a layer of 1~mm Xe gas at four different pressures (1, 5, 10 and 15~bar). The shaded band indicates the standard deviation of the computed $\alpha$ at 10~bar, providing a visual representation of the variability and uncertainty associated to the scattering angle. The scattering probability of electrons is higher at lower energies, approaching zero as the energy of the particle increases. This scattering probability also increases as a function of the gas density.

The \nexus code is run in the so-called \emph{fast-simulation mode}, meaning no scintillation photons nor ionization electrons are produced, transported, and detected. In this simulation mode, the output files only contain information about the tree of Geant4 particles (primary or secondary) in the event, and the energy deposits they leave in xenon gas for each Geant4 step.

%%%%%%%%%%
%%%%%%%%%%

In the final simulation step, we break each energy deposit into individual ionization electrons and uniformly smear along the length of the Geant4 step.  The ionization electrons are then divided into voxels of edge length 1, 2, 4, or 10–mm to capture the impact of varying detector spatial resolutions. Within each voxel, the mean position of the ionization electron cloud is used. 

This level of spatial separation among hits studied is realistic for a xenon gas detector with a tracking readout plane with sufficient granularity and coverage, and with reconstruction algorithms that are capable of unfolding the image blurring effects caused by diffusion and by the optical smearing produced in the electroluminescence gap. In particular, the existing track reconstruction in the NEXT-White detector, based on the Richardson-Lucy deconvolution algorithm \cite{Richardson:1972hli,Lucy:1974yx}, has already demonstrated a FWHM of the reconstructed Gaussian-like charge distribution of order 5~mm for both simulated and real data, independent of the drift distance, hence resolving nearby point-like energy deposits on a similar spatial separation scale \cite{NEXT:2021vzd}. 

\subsection{Event reconstruction}\label{sec:reco}
For the results presented in the next sections, we take into account the  most important limiting factors in reconstructing the kinetic energy of the most energetic electron in the event, $T_{1,\mathrm{reco}}$, and the opening angle between the electrons' initial momenta, $\cos\theta_\mathrm{reco}$. 

The non-perfect reconstruction is affected both by simulated detector effects discussed above and from our realistic classification of the various hits into $e_1$ and $e_2$ reconstructed tracks. Even assuming perfect knowledge of the decay vertex, as we do here, this task is non-trivial, given the complicated event topologies. Our approach is to use a simple reconstruction algorithm over more sophisticated methods, which will be followed up on future work. Our algorithm takes the vertex position as seed hit. We then sort all hits in the event in order of increasing distance from the vertex. The first hit (that is, the nearest hit to the vertex) is assigned to the $e_1$ group, while the second hit is assigned to the $e_2$ group. For the remaining hits, we proceed iteratively. For each hit in the event, from third to last, its distance to the last hits assigned to the $e_1$ and $e_2$ groups is computed. The hit is then assigned to the group that minimizes such distance, forming the new last hit in that group. The process ends when the farthest hit from the vertex has been classified, such that every hit is assigned to one of the two hit groups. The kinetic energy of the two decay electrons is simply the sum of the hit energy depositions in the two groups. As for the generator-level quantities, the 1$\leftrightarrow$2 electron labeling is redefined to ensure $T_{1,\mathrm{reco}}>T_{2,\mathrm{reco}}$ for all events. For the calculation of $\cos\theta_\mathrm{reco}$, we found that using the first hit associated with each track instead of fitting the first few nodes gave the best performance. This performance difference is likely due to the multiple Coulomb scattering, which quickly washes out the angular information after the first few hits. In the scenario where the closest hits were part of the same track, the angular reconstruction tends to give a small opening angle ($\cos\theta_\mathrm{reco}\sim$1). For these cases, we check if the second node in each reconstructed track produces a larger angle. If it does, we then re-assign the mis-reconstructed node to the other track, accounting for the energy re-assignment. 

%%%%%%%%%%
\begin{figure}
\centering
\includegraphics[width=\linewidth]{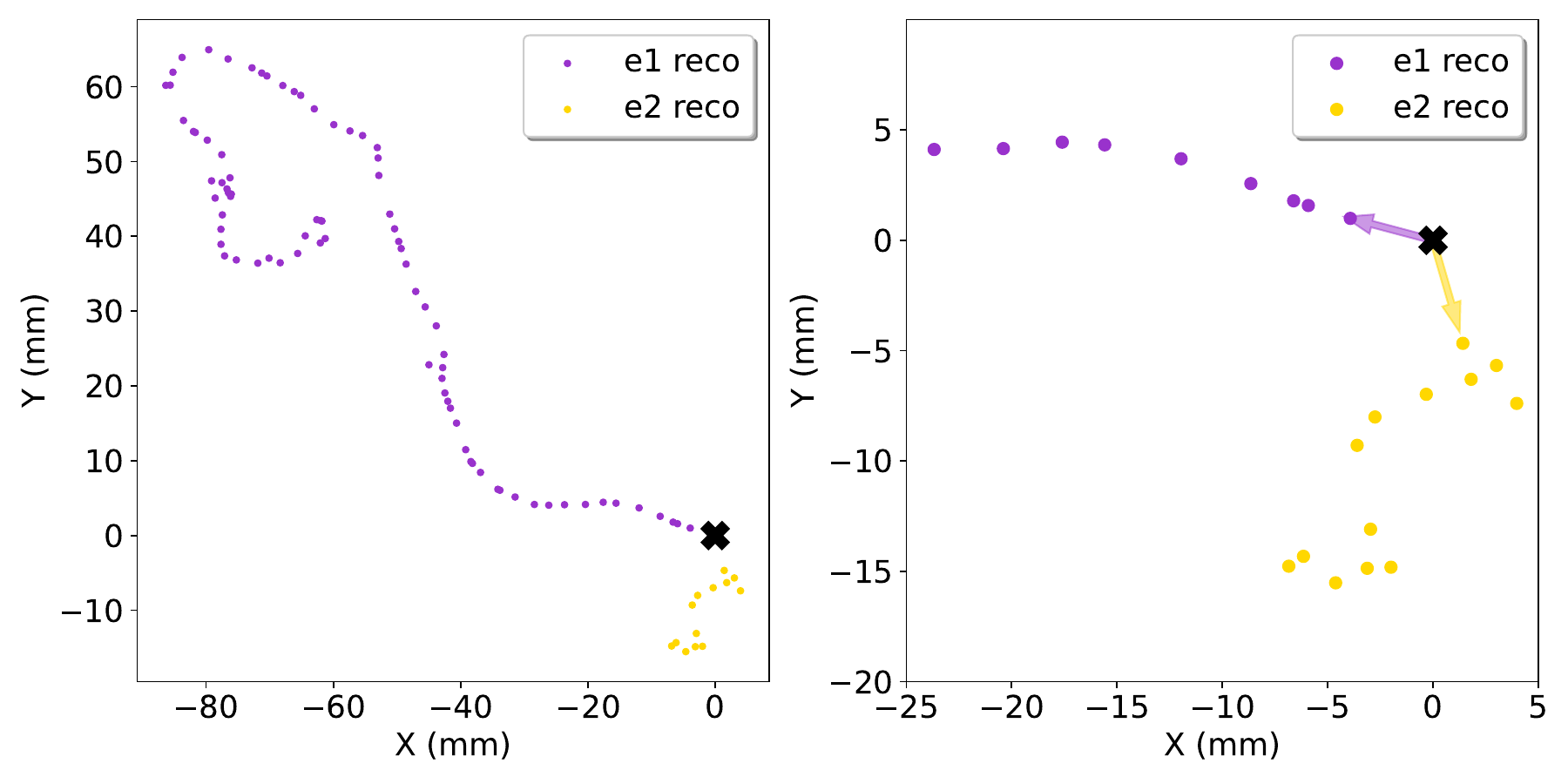}
\caption{Two-dimensional projection in the XY plane of one simulated \bbnonu~decay assuming the neutrino mass mechanism and 10~bar xenon gas pressure with a 4~mm bin size. The left panel shows the entire event, while the right panel shows a zoomed-in portion near the event vertex, indicated by a cross. The markers show the smeared and binned energy deposits, with colors indicating the assigned electron. The arrows in the right panel show the directions of the decay electrons' initial momenta projected onto the XY plane.}
\label{fig:event_display}
\end{figure}
%%%%%%%%%%

As an illustration, a two-dimensional projection of a simulated \bbnonu~decay (for the neutrino mass mechanism) at 10~bar detector pressure and a 4-mm voxelization is shown in figure~\ref{fig:event_display}. The color coding of the hits represents the reconstructed ($e_1$ or $e_2$) tracks. Clearly, the electron tracks feature fairly complicated topologies, primarily due to the large amount of multiple Coulomb scattering along the tracks in dense xenon gas. Nevertheless, the capability to reconstruct the single electron kinetic energies and initial directions is preserved.

We note that our simple reconstruction technique has been developed to address the central question of this paper of how well the energy and angular information is preserved. Future work will include improved reconstruction techniques, such as employing ML or track-building algorithms, when applied to tracks with realistic detector effects such as diffusion and assessing sensitivities to specific models.

%% file: src/reconstruction_performance.tex
\section{Event-by-Event Reconstruction Performance}
\label{sec:reconstruction_performance}
\begin{figure}
\centering
\includegraphics[width=\textwidth]{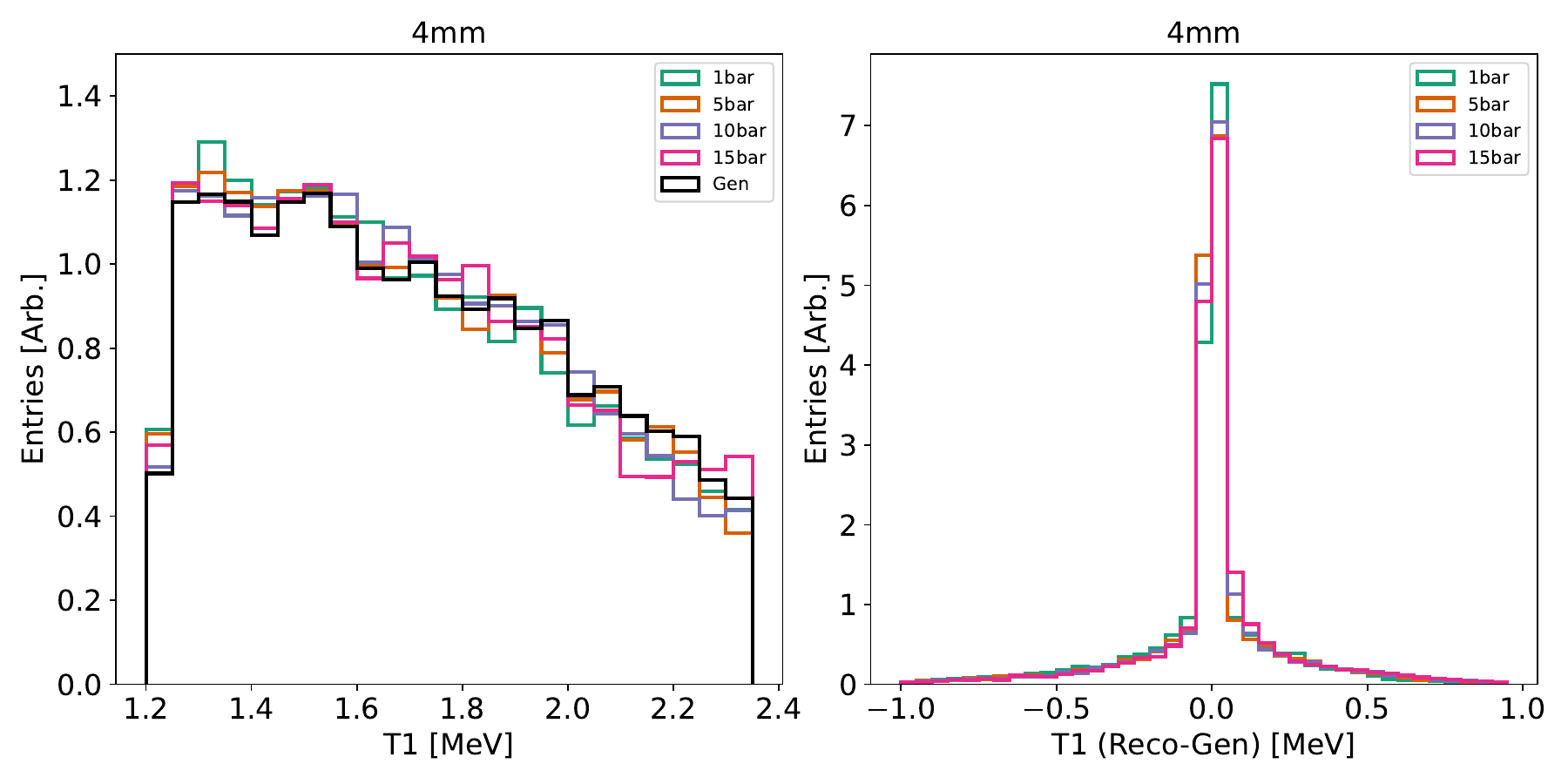} 
\caption{Left: reconstructed $T_1$ for a fixed voxelization of 4~mm for different pressures (area normalized). The black line shows the generated event distribution. Right: the distribution of reconstructed minus generated single electron kinetic energy. At this voxelization, excellent performance is achieved, with standard deviations for this difference ranging from 230 to 240~keV. There is also little variation in the performance with gas pressure.}
\label{fig:energy_reco_performance}
\end{figure}

\begin{figure}
\centering
\includegraphics[width=\textwidth]{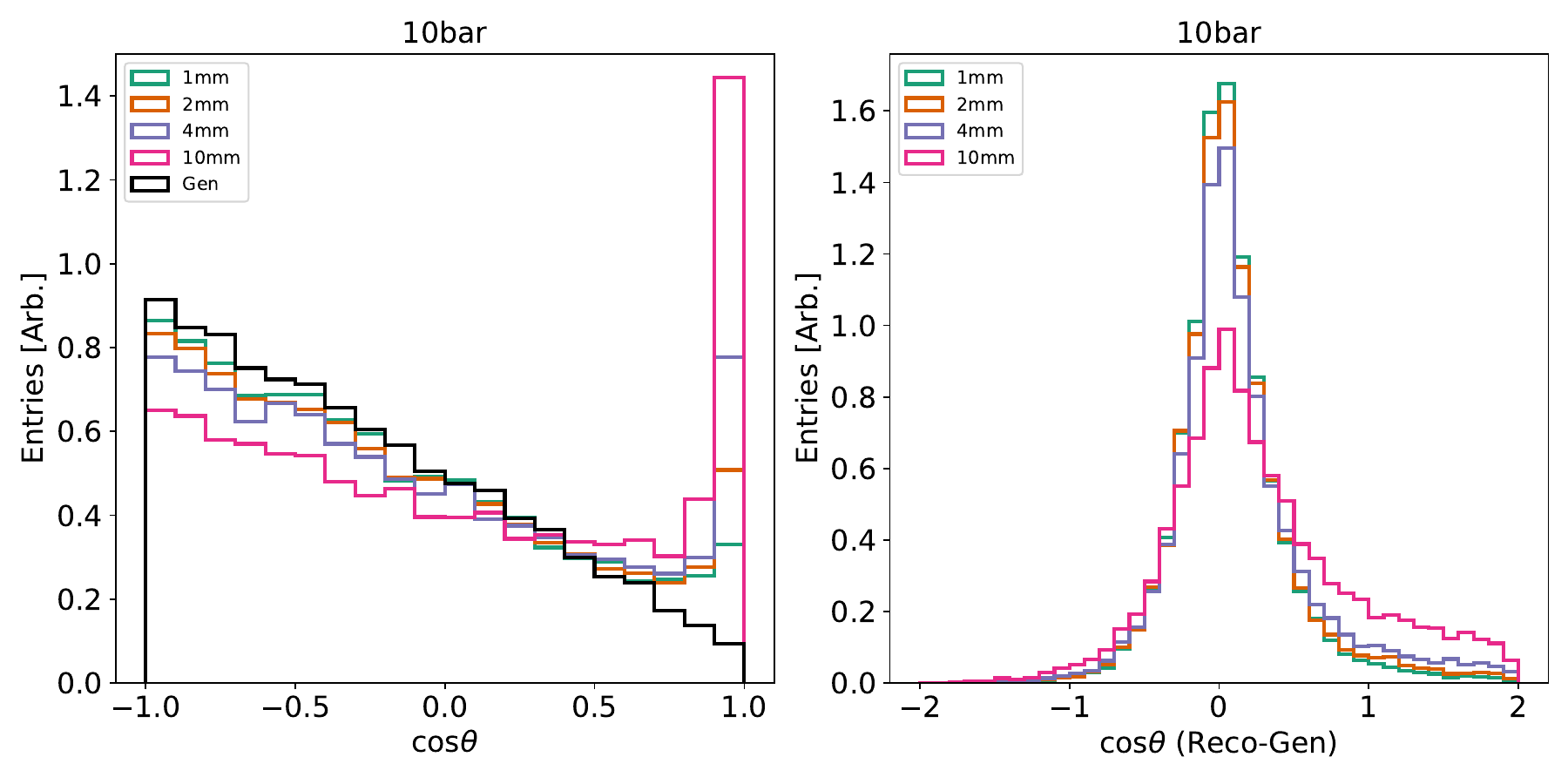} 
\caption{Left: reconstructed $\cos\theta$ for a fixed pressure of 10~bar for different voxelizations (area normalized). The black line shows the generated event distribution. Right: the distribution of reconstructed minus generated for $\cos\theta$. At this pressure, the performance starts to deviate from the generated one when the voxelization is increased. The 10-mm distribution tends towards flat, indicating that a significant amount of information in $\cos\theta$ is lost. }
\label{fig:angle_reco_performance2}
\end{figure}

Figure~\ref{fig:energy_reco_performance} shows the performance for $T_{1,\mathrm{reco}}$ reconstruction at a fixed voxelization size of 4~mm for each pressure, while figure~\ref{fig:angle_reco_performance2} shows the performance for angular reconstruction at a pressure of 10~bar for each voxelization. For the energy reconstruction, we note little dependence on pressure and voxelization. The event-by-event standard deviations of the difference between reconstructed and generated single electron energies range from 225 to 273~keV for all gas pressures and voxelizations studied. Such resolutions largely exceed the energy resolution on the $T_1+T_2$ kinetic energy sum due to the NEXT technology itself, approximately 1\% FWHM or $\sim$10~keV sigma. This justifies our approximation of not introducing any additional energy smearing effect to our analysis. For the $\cos\theta_\mathrm{reco}$ reconstruction performance, we note a stronger dependence on both the voxelization size and pressure with event-by-event reconstructed minus generated standard deviations ranging from 0.18 to 0.68. The best performance is achieved at the lowest pressure and the smallest voxelization. This is due to the improved spatial resolution and fewer scatters in a given voxel size. As can be seen in fig.~\ref{fig:angle_reco_performance2}, smaller voxels (at a fixed pressure) tend to improve reconstruction performance. By using a 10~mm voxel size at 10~bar, a significant amount of information has been lost for the $\cos\theta_\mathrm{reco}$ variable due to the multiple scattering over that length scale. 

%%%%%%%%%%
%%%%%%%%%%

%%%%%%%%%%
%%%%%%%%%%

One limitation of our hit classification algorithm is apparent in the left panel of fig.~\ref{fig:angle_reco_performance2} where there is a peak towards small opening angles near $\cos\theta_\mathrm{reco}=+1$. This peak is responsible for the asymmetric tail toward positive values in the $\cos\theta_\mathrm{reco}-\cos\theta_\mathrm{gen}$ residuals shown in the right panel of fig.~\ref{fig:angle_reco_performance2}. For a certain fraction of events, our assignment of the nearest hit to the vertex for each of the two electron tracks fails. Despite our efforts to reduce this impact, using the second-nearest node in the reconstructed track (see section~\ref{sec:reco}), it is challenging to completely remove all mis-reconstructed events. Moreover, the fraction of mis-reconstructed events with small opening angle increases with voxel size and gas pressure. Upon further inspection, many mis-reconstructed events correspond to events where there is a strong asymmetry in the energy of the electrons. The electron that carries a small fraction of the energy does not travel far, and its information can be lost within a single voxel.

%% file: src/statistical_analysis.tex
\section{Statistical Accuracy and Precision in Kinematic Observables}
\label{sec:statistical_analysis}

For each configuration described in~\ref{sec:reco}, the obtained datasets are then divided into several subsamples, each containing a fixed number $n<N$ of detected \bbnonu\ signal events (with $N=6000$), and constituting the toy datasets that allow a statistical analysis as a function of $n$. The number of events contained in each dataset varies in the $[5, 100]$ range. The lower limit $n=5$ is motivated by having a minimal amount of shape information on the reconstructed kinematic variables. The upper limit $n=100$ corresponds to a favorable scenario based on current half-life limits and future experimental projections. In particular, $n=100$ \bbnonu events would be detected in a fully efficient xenon detector with 1.2~tonnes of \Xe{136} fiducial mass, accumulating 10 years of exposure, and assuming a \Xe{136} half-life as short as permitted by the latest results from KamLAND-Zen, $\halflife>3.8\times 10^{26}$ yr at 90\% CL \cite{KamLAND-Zen:2024eml}. In the following, we study how well we can reconstruct the following average quantities:
\begin{equation}
\label{eq:acc}
\overline{X} = \frac{1}{n}\sum_{i=1}^n X_{i}\,,
\end{equation}
where $X = \{T_{1}, \cos\theta \}$ and $n = \{5, 10, ..., 100\}$. In particular, we evaluate two separate performance indicators: accuracy and precision. The accuracy, or bias, refers to how well the energy or angle reconstructed average values reproduce the generated average values, in the limit of very large signal sample statistics ($n\to\infty$). The precision, or resolution, refers instead to the variations from one signal dataset to another, in the realistic finite $n$ case, in reconstructing the average values. We are particularly concerned with precision. While biases can largely be corrected for, although some model dependence may be introduced in the process, resolution effects constitute an irreducible uncertainty in reconstructing the energy and angle average values. Numerically, we estimate the accuracy as the difference between reconstructed and generated average values over the full datasets with $N$ events. For a fixed number of signal events $n<N$, we estimate the precision as the standard deviation of the average values obtained from the $N/n$ toy datasets. We also compute the error on the estimation of the mean reconstructed values by dividing the precision by the square root of the number of toys. What follows presents the dependence of the accuracy and precision on the average $\cos\theta$ and $T_{1}$ values as a function of different factors such as the signal sample size $n$, detector pressure and readout granularity, and different nuclear and atomic physics assumptions.

\subsection{Impact of signal sample size}
\begin{figure}
\centering
\includegraphics[width=\textwidth]{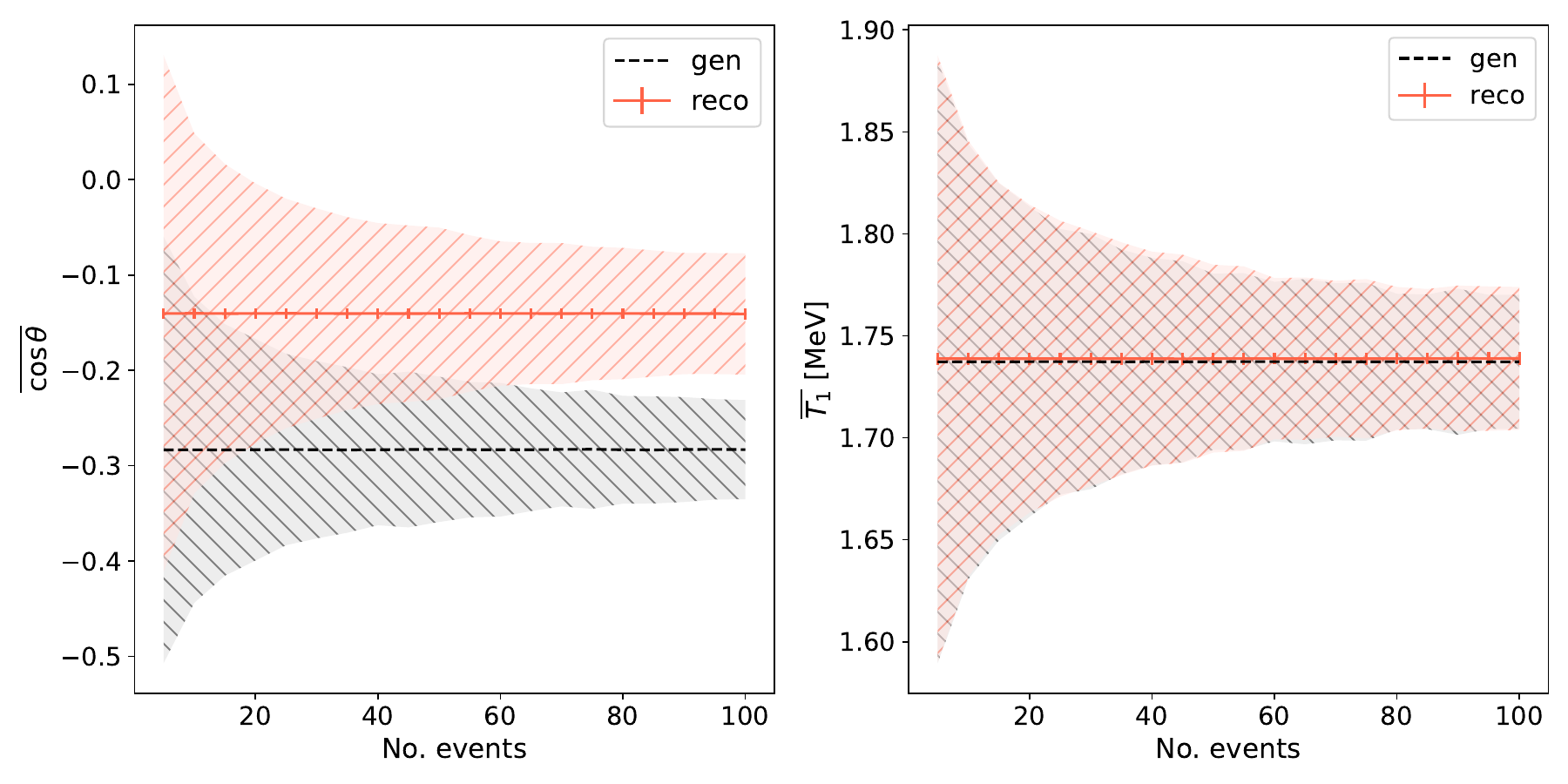}
\caption{Mean (central lines), standard deviation (band widths) and error on the estimation of the mean (error bars) of the reconstructed and generated average values $\overline{\cos\theta}$ (left) and $\overline{T_{1}}$ (right), as a function of the number of signal events in each toy dataset. The plots assume a gas pressure of 10~bar and a detector voxelization of 4~mm as a reference.}
\label{fig:both_nsig}
\end{figure}

In the left panel of figure~\ref{fig:both_nsig}, the mean and standard deviation for the opening angle average values from each toy dataset are shown as a function of the number of signal events, for a pressure of 10~bar and a voxelization of 4~mm. The reconstructed and generated average value bands are both shown, as well as the error in the estimation of the mean. From the difference between the two mean values, shown as the horizontal central lines, a $\overline{\cos\theta}$ accuracy of +0.14 is expected in these experimental conditions, where the plus sign indicates that detector and reconstruction effects biases towards larger $\overline{\cos\theta}$ values, compared to generated ones. The standard deviations are shown as band widths. As expected, the standard deviation decreases with increasing $n$ for both the reconstructed and generated average values. In particular, for the reconstructed $\overline{\cos\theta}$, the precision varies from 0.27 to 0.06 in the $[5, 100]$ signal event range, and it is therefore comparable in size with the reconstruction accuracy.

%%%%%%%%%%
%%%%%%%%%%

The corresponding results for the single electron energy average values are shown in the right panel of fig.~\ref{fig:both_nsig}. In this case, the accuracy is much better than the precision, with a $\overline{T_1}$ bias of only 1.5~keV, or 0.09\% of the \nudobe-generated single electron energy average value of 1.74~MeV for the standard neutrino mass mechanism. In this case, the precision improves from 150 to 40~keV over the $[5, 100]$ event range. 

In general, we expect that the width of the reconstructed bands in fig.~\ref{fig:both_nsig} depends on two factors: the number of events detected and the accuracy of the reconstruction. For both reconstructed variables, the error of the estimation of the mean yields very small values: on average, a value of 0.005 for the reconstructed opening angle and of 3 keV for the single electron energy.  By comparing the reconstructed bands with the generated ones, also shown in the figure and having a similar width, we conclude that the precision is mostly dictated by the finite signal statistics. As mentioned above, the most important reconstruction limitation turns out to be the bias in the angular reconstruction, which varies with the experimental conditions, being worse at higher pressure and coarser tracking granularity.

\subsection{Impact of detector pressure and readout granularity}
\begin{figure}
\centering
\includegraphics[width=\textwidth]{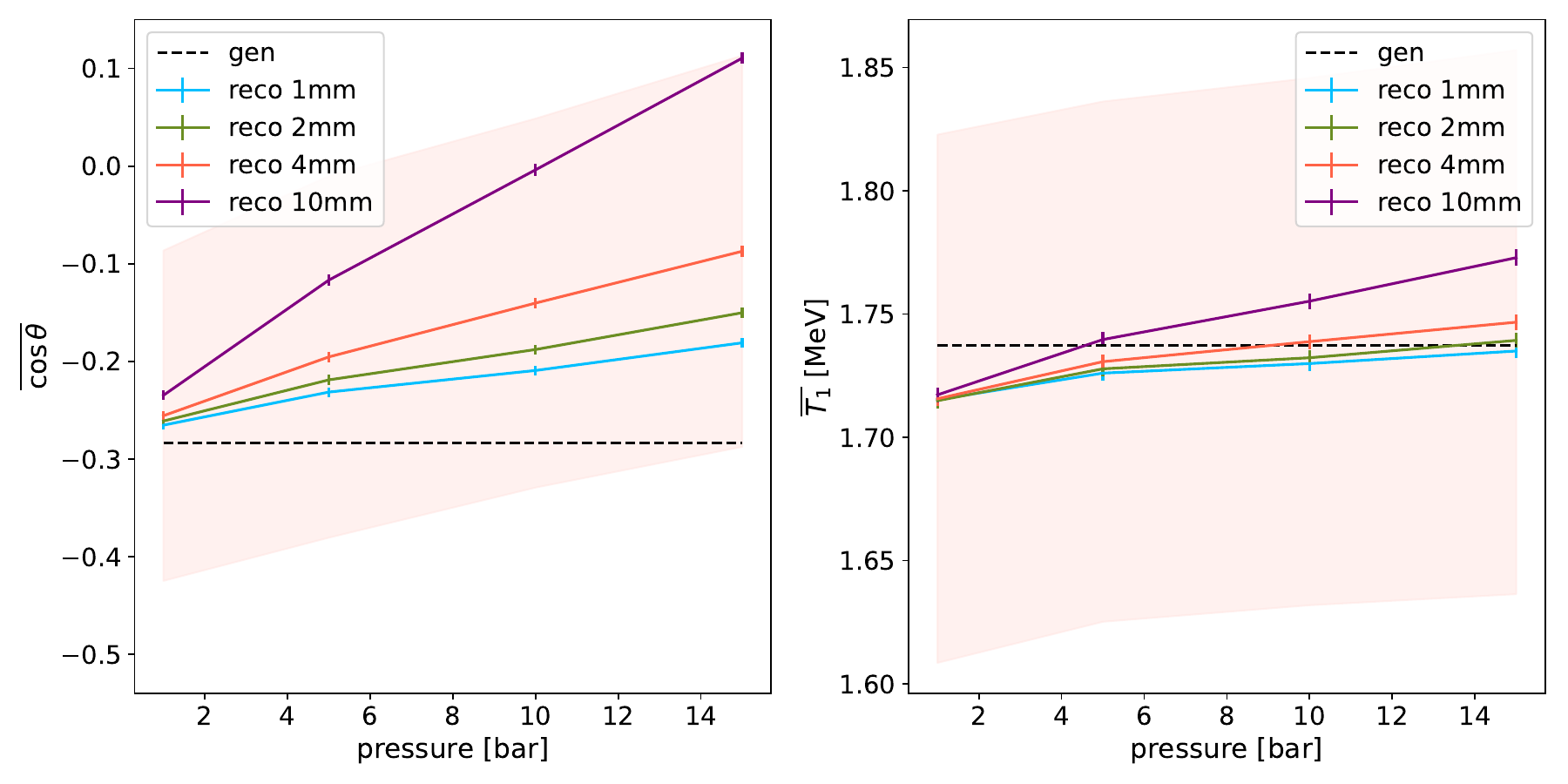}
\caption{Mean (central lines), standard deviation (band widths) and error on the estimation of the mean (error bars) of the reconstructed and generated average values $\overline{\cos\theta}$ (left) and $\overline{T_{1}}$ (right) for different voxel sizes, as a function of detector pressure. The plots assume a number $n=10$ of signal events in each toy dataset as a reference. For figure clarity, band widths for reconstructed average values and 4~mm voxelization only are shown.}
\label{fig:both_pressure}
\end{figure}

Figure~\ref{fig:both_pressure} illustrates the dependence of reconstructed and generated average values with gas pressure, for different voxelization sizes. As in fig.~\ref{fig:both_nsig}, the difference between the reconstructed and generated central lines yields the accuracy. The \nudobe-generated average values cannot depend on detector pressure, while reconstructed average values can. In particular, for $\overline{\cos\theta}$, the difference between the generated and reconstructed average values increases with pressure, from +0.03 to +0.19 in the 1--15~bar range and for a fixed 4~mm voxelization. The absolute error on the estimation of the mean of the $\overline{\cos\theta}$ is on average of 0.005, slightly increasing with pressure. For figure clarity, the standard deviation for the generation-level observables is not shown for any detector pressure, nor readout granularity. On the other hand, the standard deviation for the reconstruction-level observables is shown for all detector pressures but only for readout granularities of 4~mm. The standard deviation of the reconstructed $\overline{\cos\theta}$ value from the various toy datasets at 1~bar is very similar to the \nudobe-generated spread, while it is somewhat larger for larger pressures. In general, the pressure dependence for the bias is much stronger than for the resolution, particularly for coarser voxelizations.

%%%%%%%%%%
%%%%%%%%%%

For the other kinematic variable, $\overline{T_{1}}$, there is also a small pressure-dependent pattern for the bias and the resolution. The bias slightly increases with increasing pressure at coarser voxelizations. Mean biases of a few tens of keV at most, and resolutions of order 100~keV, are obtained over the entire pressure range, for a fixed 4~mm voxelization and $n=10$ signal events. In this case, the error on the average reconstructed variable remains at a constant value of 3 keV throughout the entire pressure [1,15] bar range.

\subsection{Impact of nuclear and atomic physics uncertainties}
Finally, we compare the variations in the average kinematic values arising from variations in the nuclear, hadronic and atomic physics assumptions, all within \nudobe. Overall, we explore 12 different configurations at generation level: three possible assumptions for the long-range part of the NMEs, times two different short-range NME contribution assumptions, times two different PSF assumptions. The long-range NME variations studied include calculations using the \emph{interacting boson model} (IBM2) \cite{Deppisch:2020ztt} and the \emph{quasi random-phase space approximation} (QRPA)  \cite{Hyvarinen:2015bda} in addition to our reference \emph{shell model} (SM) \cite{Menendez:2017fdf} calculation. We study the impact of recently identified short-range contributions to NMEs arising from hard neutrino-exchange processes \cite{Cirigliano:2018hja,Cirigliano:2019vdj}, compared to our nominal assumption of negligible short-range contributions. Finally, we explore the impact of an alternative PSF scheme, one where PSFs are computed exactly assuming a point-like nucleus, unlike our nominal calculation where PSFs are calculated from approximate wave functions in a uniform nuclear charge distribution \cite{Stefanik:2015twa}. 

Out of the 12 \nudobe variations explored, we find that the generation with a point-like nucleus PSF assumption has a small difference with respect to the calculation using approximate wavefunctions. In this case, the \nudobe-generated $\overline{\cos\theta}$ value is about 0.01 units lower than the $-0.28$ value obtained with our nominal analysis. While appreciable, this bias is in any case sub-dominant compared to the reconstruction angular biases (+0.14 at 10~bar pressure and 4~mm voxelization). For $\overline{T_{1}}$, generator-level variations from the various \nudobe assumptions yield variations within $\sim 1$~keV, comparable to reconstruction biases and much smaller than $\overline{T_{1}}$ resolutions for realistic numbers of signal events detected.

%% file: src/conclusions.tex
\section{Summary and Conclusions}
\label{sec:conclusions}
In this study, we demonstrate the statistical accuracy and precision in measuring the electron kinematic observables in a future NEXT detector. Such observables could be used to discriminate among neutrinoless double beta (\bbnonu) decay contributions generated by different lepton number violating (LNV) mechanisms. Specifically, the distributions of the opening angle between the two emitted electrons ($\cos\theta$) and their individual kinetic energies ($T_1$ and $T_2$) vary significantly depending on the dominant LNV contribution to the \bbnonu-decay amplitude. Using realistic simulations, we show here that a xenon gas time projection chamber complemented with vertex-tagging capabilities and sufficient readout resolution ---\thinspace either deriving from advanced track reconstruction or from barium tagging \thinspace--- can preserve $\cos\theta$ and $T_{1,2}$ information.

Our results account for the most relevant detection and reconstruction effects, particularly the Coulomb elastic scattering of the decay electrons in dense xenon gas, and the imperfect capability to correctly associate the energy deposits of the complex event topologies to the corresponding electron tracks. We also consider gas pressure, readout granularity, and the impact of nuclear and atomic physics uncertainties. 

For realistic detector pressure and spatial resolution conditions of 10~bar and 4~mm, respectively, we find that the average $\cos\theta$ and $T_1$ values can be reconstructed with a precision of $+0.19$ and 110~keV, respectively, assuming that only 10 \bbnonu events are detected.

Such signal statistics are reachable in a tonne-scale xenon gas detector, provided we discover the {\bbnonu} decay of \Xe{136} at a rate that is not much slower than current bounds \cite{KamLAND-Zen:2024eml}. If the \Xe{136} {\bbnonu} half-life turns out to be even longer, our results may serve to establish a minimal xenon gas detector mass to be able to discriminate different \bbnonu-decay mechanisms.

Detailed studies on whether the precisions achieved in this study in reconstructing kinematic variables are sufficient for physics model discrimination will be discussed in future papers. One difficulty is to scan in a systematic and unbiased way the large multidimensional free parameter space defining the alternative \bbnonu mechanisms of the type shown in fig.~\ref{fig:model_kinematics}. However, we can already make some qualitative statements. In general, the significance in discriminating among two alternative models using some kinematic variable (for example, $\overline{\cos\theta}$ or $\overline{T_1}$, as studied here) is directly proportional to the difference among the models’ preferred values for such variable, and inversely proportional to the quadrature sum of the precision achievable on that variable for the two models. On the one hand, we have found that we obtain similar precisions in reconstructing kinematic variables for alternative models compared to the standard neutrino mass mechanism. On the other, we have also observed that kinematic information gathered from angular information is mostly uncorrelated from the one obtained from electron kinetic energy sharing, and hence the capability of reconstructing both kinematic quantities at the same time (as demonstrated here) is a definite asset. Putting all this together, we believe that prospects for discriminating $\overline{\cos\theta}\gtrsim 0$ and/or $\overline{T_1}\gtrsim 2.0$~MeV models specifically are bright, provided that a sufficient number of \bbnonu events can be detected.

%% file: src/ack.tex
The NEXT Collaboration acknowledges support from the following agencies and institutions: the European Research Council (ERC) under Grant Agreement No.\ 951281-BOLD; the European Union’s Framework Programme for Research and Innovation Horizon 2020 (2014–2020) under Grant Agreement No.\ 860881-HIDDeN; the MCIN/AEI of Spain and ERDF A way of making Europe under grants PID2021-125475NB and RTI2018-095979, and the Severo Ochoa and Mar\'ia de Maeztu Program grants CEX2023-001292-S, CEX2023-001318-M and CEX2018-000867-S; the Generalitat Valenciana of Spain under grants PROMETEO/2021/087 and CISEJI/2023/27; the Department of Education of the Basque Government of Spain under the predoctoral training program non-doctoral research personnel; the Spanish la Caixa Foundation (ID 100010434) under fellowship code LCF/BQ/PI22/11910019; the Portuguese FCT under project UID/FIS/04559/2020 to fund the activities of LIBPhys-UC; the Israel Science Foundation (ISF) under grant 1223/21; the Pazy Foundation (Israel) under grants 310/22, 315/19 and 465; the US Department of Energy under contracts number DE-AC02-06CH11357 (Argonne National Laboratory), DE-AC02-07CH11359 (Fermi National Accelerator Laboratory), DE-FG02-13ER42020 (Texas A\&M), DE-SC0019054 (Texas Arlington) and DE-SC0019223 (Texas Arlington); the US National Science Foundation under award number NSF CHE 2004111; the Robert A Welch Foundation under award number Y-2031-20200401. This research was also done using services provided by the OSG Consortium \cite{osg1,osg2,osg3,osg4}, which is supported by the National Science Foundation awards \#2030508 and \#1836650. F.~Pompa acknowledges support from the Generalitat Valenciana of Spain through the “plan GenT” programme (CIDEGENT/2018/019). JM-A acknowledges support from the Ram\'on y Cajal program, grant RYC2021‐033265‐I, funded by MCIN/AEI/10.13039/501100011033 and by the EU \guillemotleft NextGenerationEU\guillemotright/PRTR.
Finally, we are grateful to the Laboratorio Subterr\'aneo de Canfranc for hosting and supporting the NEXT experiment.